\documentclass[12pt,a4paper]{article}
\usepackage{physics}
\usepackage{amsfonts}
\usepackage{amsmath}
\usepackage{amssymb}
\usepackage{amsthm}
\usepackage{graphicx}
\usepackage{jheppub}
\usepackage{url}
\usepackage{xcolor}
\usepackage[utf8]{inputenc}
\usepackage[T1]{fontenc}
\usepackage{lmodern}
\usepackage[section]{placeins}
\definecolor{refkey}{gray}{0.45}
\definecolor{labelkey}{RGB}{155,48,48}
\usepackage[section]{placeins}
\usepackage[utf8]{inputenc}
\usepackage[T1]{fontenc}
\usepackage{framed}
\hypersetup{
  colorlinks=true,
  citecolor=purple,
  linkcolor=blue,
  urlcolor=violet
 }
 
 \newcommand{\mi}{\mathrm{i}}

\newcommand{\bi}{\begin{itemize}}
\newcommand{\ei}{\end{itemize}}
\newcommand{\non}{\nonumber}
\newcommand{\bea}{\begin{eqnarray}}
\newcommand{\eea}{\end{eqnarray}}
\newcommand{\be}{\begin{equation}}
\newcommand{\ee}{\end{equation}}
\newcommand{\ben}{\begin{eqnarray*}}
\newcommand{\een}{\end{eqnarray*}}
\newcommand{\bem}{\begin{pmatrix}}
\newcommand{\eem}{\end{pmatrix}}
\newcommand{\bl}{\begin{align}}
\newcommand{\el}{\end{align}}
\newcommand{\beg}{\begin{gather}}
\newcommand{\eeg}{\end{gather}}

\newcommand{\cH}{\mathcal{H}}

\newcommand{\bM}{\ensuremath{\mathbb{M}}}

\newcommand{\IH}{\mathbb{H}}

\renewcommand{\b}{\beta}

\newcommand{\e}{\epsilon}

\renewcommand{\k}{\kappa}             

\newcommand{\m}{\mu}

\newcommand{\p}{\pi}
\renewcommand{\r}{\rho}                                     
                                   
\renewcommand{\t}{\tau}

\newcommand{\TrH[1]}{ {\raise -.5em
                      \hbox{$\buildrel {\textstyle  {\rm Tr } }\over
{\scriptscriptstyle \cH _ {#1}}$}~}}
\newcommand{\res[1]}{{\raise -.5em 
\hbox{$\buildrel{\textstyle{\rm Res}}\over {\scriptscriptstyle {#1}}$}}}
\newcommand{\tends[1]}{{\raise -.5em 
\hbox{$\buildrel{\longrightarrow}\over {\scriptscriptstyle {#1}}$}}}

\newcommand{\pa}{\partial}

\renewcommand{\Im}{\mbox{Im}}
\renewcommand{\Re}{\mbox{Re}}

\def\dbend{\lower3.5pt\hbox{\manual\.}}

\def\IL{\relax{\rm I\kern-.18em L}}
\def\IH{\relax{\rm I\kern-.18em H}}
\def\rlx{\relax\leavevmode}

\def\ZZ{\rlx\leavevmode\ifmmode\mathchoice{\hbox{\cmss Z\kern-.4em Z}}
 {\hbox{\cmss Z\kern-.4em Z}}{\lower.9pt\hbox{\cmsss Z\kern-.36em Z}}
 {\lower1.2pt\hbox{\cmsss Z\kern-.36em Z}}\else{\cmss Z\kern-.4em
 Z}\fi}

\title{\center\textmd{Quantum Entanglement on Black Hole Horizons \\in 
String Theory and  Holography}}

\preprint{}
\author{ \center Atish  Dabholkar and} \author{Upamanyu Moitra}
\affiliation{\begin{center}
The Abdus Salam International Centre for Theoretical Physics\\ Strada Costiera 11, Trieste 34151, Italy
\end{center}  }

\abstract{We compute the exact one-loop partition function of $\mathbb{Z}_N$ orbifolds of Euclidean BTZ black hole with the aim to compute the entanglement entropy of the black hole horizon in string theory as a function of the mass and spin of the black hole and the $\mathrm{AdS}_3$ radius. We analyze the  tachyonic  contribution  to the modular integrand for the partition function known for odd integers $N>1$ and show that it admits an analytic continuation resulting in a finite answer for the modular integral in the physical region $0< N \leq 1$. We discuss the flat space limit and  the relevance of this computation for quantum gravity near black hole horizons and holography in relation to the thermal entropy. 
}


\makeatletter
\gdef\@fpheader{}
\makeatother
\begin{document}

\maketitle

\section{Introduction}

An approach for defining and computing entanglement entropy in string theory was proposed in \cite{Dabholkar:1994ai,Dabholkar:2001if,  Dabholkar:2022mxo} using partition functions $Z(N)$ of orbifolds of string theory for any odd integer $N$ as the starting data \cite{Dabholkar:1994ai,Dabholkar:2001if,  Dabholkar:2022mxo}.  
In the simplest example of the ten-dimensional Type-II string theory, the relevant 
orbifolds are  $\bM_8 \times \mathbb{R}^2/\mathbb{Z}_N$  where $\bM_{8}$ is $8$-dimensional flat space and $\mathbb{R}^2$ is the Euclidean Rindler plane \cite{Dabholkar:1994ai}. The partition function $Z(N)$ can be identified with the logarithm of $\Tr(\rho^{1/N})$ where $\rho$ is the reduced density matrix for the right Rindler wedge \cite{Dabholkar:1994ai, Witten:2018xfj}. If one can find a suitable analytic interpolation to the physical region $0 <  N \leq  1$ then one can indirectly define an analog of von Neumann entropy and possibly R\'enyi entropy $S_r(N)$ as a function of $N$. 
The orbifold method can be viewed as a stringy generalization of the replica method in quantum field theory for computing the entanglement entropy associated with the entangling surface at the Rindler horizon. 

It is well-known that entanglement entropy in quantum field theory  is ultraviolet-divergent and is proportional to the area of the entangling surface. Since modular integrals for partition functions in string theory have a natural ultraviolet cutoff, one expects that the corresponding quantity in string theory would be free of the ultraviolet divergence. One may therefore hope to obtain a well-defined finite answer in string theory.  However, as noted in \cite{Dabholkar:1994ai,  Dabholkar:2023ows},  one has to contend with infrared tachyonic  divergences of the integral for $Z(N)$ to be able to implement this idea. 

It was shown in \cite{Dabholkar:2023ows} that it is possible to resum the tachyonic divergences and analytically continue the resulting answer to the physical region $0< N \leq  1$ where the integral is no longer divergent. Subsequently, it was shown in \cite{Dabholkar:2023yqc}  that this method generalizes beyond the ten-dimensional Type-II theory for several string compactifications on orbifolds with lower supersymmetry. The tachyonic resummation and analytic continuation works even when there are additional tachyons from doubly-twisted sectors in some of these examples. 
With these results, the orbifold method seems to offer a viable route for defining and computing a notion of entanglement entropy in quantum gravity that is finite both in the ultraviolet and the infrared. For earlier attempts,  see \cite{He:2014gva}.  As emphasized in \cite{Dabholkar:2022mxo,  Dabholkar:2023ows}, finiteness of entanglement entropy has important implications in quantum gravity, for example, for the information paradox.

 It is of interest to generalize these considerations near the horizon of a black hole with finite mass $M$. One can then examine if it is possible to define a notion of finite entanglement entropy as a function of two variables $S_r(N, M)$ measuring the entanglement between the inside and outside of the black hole.   In the limit when $M$ tends to infinity, the near-horizon geometry of a black hole with a bifurcate horizon can be approximated by Rindler spacetime with Rindler horizon. The left  and  right exterior regions of the two-sided black hole geometry correspond to the left and right Rindler wedges respectively. The Hartle-Hawking  vacuum $|\Omega\rangle$ of the black hole corresponds to the Minkowski vacuum for the Rindler geometry describing the near-horizon region. 
 In this limit, $S_r(N, M)$ should reduce to $S_r(N)$. 
 
 In fact, it is worth considering a further generalization by embedding the black hole in a holographic context and consider an eternal Kerr black hole in Anti-de Sitter ($\mathrm{AdS}$) spacetime.   
In this case, the starting data would be partition functions $Z(N, M, J, L)$  of  $\mathbb{Z}_N$ orbifolds of the Euclidean black hole horizon depending on four variables, where $J$ is the spin of the black hole and $L$ is the $\mathrm{AdS}$ radius. One can then hope to obtain the R\'enyi entropy as a function of four variables $S_r(N, M, J, L)$.  An additional advantage of embedding the problem in holography is that one can then compare the bulk gravity with the boundary CFT to gain a better physical understanding of the `algebra of observables' in the bulk. 

To implement these ideas,  it is necessary to be able to explicitly construct the orbifold partition function, which requires a description of the Euclidean horizon geometry in terms of an exact world-sheet conformal field theory. 
For this purpose, we consider the near-horizon geometry the F1-NS5 system in Type-II theory compactified on $T^4 \times S^1$. A spinning black hole in the noncompact five-dimensional spacetime corresponds to the Ba\~nados-Teitelboim-Zanelli (BTZ) black hole \cite{Banados:1992wn} with mass $M$ and spin $J$ in asymptotically $\mathrm{AdS}_3$ spacetime.  A BTZ black hole in the NS-NS background admits an exact conformal field theory description and   provides  a suitable starting point for constructing the $\mathbb{Z}_N$ orbifolds. The bifurcate horizon of the black hole in the Euclidean continuation corresponds to a fixed point of the orbifold symmetry and the spatial horizon corresponds to the entangling surface. 

The conformal field theory, even though exact, is nevertheless a nontrivial interacting quantum field theory. Somewhat surprisingly, we find that it is possible to construct the orbifold partition function quite explicitly. The multiplicities and masses of tachyons can be then be easily read off.  We find that, just as in flat space, the tachyonic contribution to the partition function admits an analytic continuation which is finite in the physical domain $0 < N \leq 1$ with a nontrivial dependence on $\k$, $M$ and $J$. These considerations can be easily generalized to other compactifications with an $\mathrm{AdS}_3$ factor, for example, with  $T^4$ above replaced by $\mathbf{K3}$. 

The paper is organized as follows. In \S\ref{sec: BTZ} we review relevant facts about the BTZ black hole \cite{Banados:1992wn} in the near-horizon $\mathrm{AdS}_3$ geometry of the F1-NS5 system. In \S\ref{sec:Partition} we construct the exact partition functions of the $Z_N$ orbifolds of the $\mathrm{AdS}_3 \times S^3 \times T^4$ spacetime as functions of $N, M, J, L$. The orbifold action leaves fixed the Euclidean horizon of the BTZ black hole which can thus be identified with the entangling surface separating the inside and the outside of the black hole. In \S\ref{sec:Ent} we resum the tachyonic contributions to the partition function and show that the analytic continuation of the resulting sum is free of tachyonic divergences in the physical domain $0<N \leq 1$. We conclude with a discussion of the flat space limit and the relation between entanglement entropy and the thermal entropy in the context of $\mathrm{AdS}_3/\mathrm{CFT}_2$ holography.

\section{Strings on Black Hole Backgrounds \label{sec: BTZ}}

Consider Type-IIB string theory on $\mathbb{R}^{1,4} \times T^4 \times S^1$  background  where $\mathbb{R}^{1,4}$ is a Lorentzian spacetime, $T^4$ is a 4-torus and $S^1$ is a circle.  This compactification results in a five-dimensional theory  with $32$ supercharges. We consider $Q_1$ F1-strings wrapping the $S^1$ and $Q_5$ NS5-branes wrapping $T^4 \times S^1$ corresponding to a point-like charged state in five dimensions. We are interested in studying the entanglement entropy across the horizon of a black hole with these charges with mass $M$ and spin $J$ in one of the planes  of the spatial $\mathbb{R}^4$. The black hole can be regarded as an excitation of the F1-NS5 system. To focus on the entangling surface which coincides with the black hole horizon, we consider string theory in the near-horizon geometry. 

The near-horizon geometry of the F1-NS5 system is $\mathrm{AdS}_3 \times S^3 \times T^4$ with NS-NS background fields. The five-dimensional black hole corresponds to a BTZ black hole in the $\mathrm{AdS}_3$. In this section we first recall the description of both the Lorentzian and Euclidean black hole in the supergavity limit and then as an exact  superconformal field theory (SCFT) to prepare for the orbifold construction. 
For completeness,  we also discuss the ${\rm SL}(2, \mathbb{Z})$ family of solutions related to the BTZ black hole. More details can be found in the review \cite{Kraus:2006wn} and references therein.

\subsection[Strings on $\mathrm{AdS}_3 \times S^3 \times T^4$]{\boldmath Strings on $\mathrm{AdS}_3 \times S^3 \times T^4$ \label{AdS3-CFT}}

We first consider the more familiar D1-D5 system \cite{Strominger:1996sh,  Breckenridge:1996is} with  $Q_1$ D1 branes and $Q_5$ D5-branes in Type-IIB compactified on $T^4$. The near-horizon  
string-frame metric \cite{Polchinski:1998rr} is
\begin{equation}
\dd s^2 = \frac{r^2}{r_1 r_5} \eta_{\mu \nu} \dd x^\mu \dd x^\nu + \frac{r_1 r_5}{r^2} \dd r^2 + r_1 r_5 \dd \Omega_3^2 + \frac{r_1}{r_5} \delta_{ij} \dd x^i \dd x^j,  \label{adbs31}
\end{equation}
where $\eta_{\mu \nu}$ is the two-dimensional Minkowski metric,  $\dd \Omega_{3}^2$ is the line-element on the unit 3-sphere and the coordinates $x^i$ parametrize the torus $T^4$, with volume $(2\pi)^4 V_4$.
The dilaton is given by
\begin{equation}
e^{-2\Phi} = \frac{r_5^2}{r_1^2}  \, .\label{dil1} 
\end{equation} 
The parameters $r_{1,5}$ are related\footnote{Factors of string coupling $g$ better suited for our purposes here differ slightly compared to \cite{Polchinski:1998rr}.}
to the D-brane charges $Q_{1,5}$,  asymptotic string coupling $g$ and  the string length $\ell_s \equiv \sqrt{\alpha'}$ as
\begin{align}
r_1^2 &=  \frac{g^2 Q_1 \ell_s^6}{V_4}, \label{dr1} \\
r_5^2 &= Q_5 \ell_s^2. \label{dr2}
\end{align}
The metric \eqref{adbs31} evidently describes the product spacetime $\mathrm{AdS}_3 \times S^3 \times T^4$. The radius $L$ of $S^3$  is given by $L^2 = r_1 r_5$ and equals the  radius of curvature of $\mathrm{AdS}_3$.

The F1-NS5 background of our interest \cite{Callan:1991ky,  Tseytlin:1996as} with  $Q_1$ fundamental  strings and $Q_5$ NS5-branes  can be obtained by an
 $S$-duality transformation \cite{Maldacena:1998bw,  Giveon:1998ns}.  Under  $S$-duality, the dilaton and string-frame metric transform as
\begin{equation}
\Phi \to -\Phi,   \quad \dd{s}^2 \to e^{-\Phi} \dd{s}^2.  \label{sdutra}
\end{equation}
The near-horizon string frame metric for the F1-NS5 system is then given by
\begin{equation}
\dd s^2 = \kappa \left( \frac{r^2}{\ell_s^2} \eta_{\mu \nu} \dd x^\mu \dd x^\nu + \frac{\ell_s^2}{r^2} \dd r^2 \right) +  \kappa \ell_s^2 \dd \Omega_3^2 +  \delta_{ij} \dd x^i \dd x^j, \label{metns2}
\end{equation}
after a suitable rescaling of the two-dimensional $x^\mu$  and the toroidal $x^i$ coordinates, where $\kappa \equiv Q_5$ is the NS5-brane charge. Correspondingly, there are $\kappa$ units of quantized magnetic flux of the NS-NS three-form on the $S^3$.  The $S^3$ radius $L$ is given by $L^2 = \ell_s^2 \kappa $ and equals the $\mathrm{AdS}_3$ radius of curvature. Note that in this frame, the geometry appears to depend on only one of the charges of the original theory, namely $Q_5$, and one may wonder about the role of $Q_1$. It is known that $Q_1$ appears as the limit on the spectrally flowed representations allowed in theory. See, for example, \cite{Dabholkar:2007ey} for a summary.

An appealing feature of the F1-NS5 system is that this background admits an exact conformal field theory description. The $S^3$ factor is the group manifold of $\mathrm{SU}(2)$ with flux, which can be described as an $\mathrm{SU}(2)$ Wess-Zumino-Witten (WZW) model. Lorentzian $\mathrm{AdS}_3$ is the coset manifold $\mathrm{SO} (2,2) / \mathrm{SO} (2,1)$.  Ignoring global identifications,  one has
\begin{equation}
\mathrm{SO} (2,2) \cong \mathrm{SL} (2,  \mathbb{R} ) \times \mathrm{SL} (2,  \mathbb{R} ),  \quad \mathrm{SO} (2,1) \cong  \mathrm{SL} (2,  \mathbb{R} ). \label{so22e}
\end{equation}
We are interested in the Euclidean $\mathrm{AdS}_3$ which is the coset manifold $\mathrm{SO} (3,1) / \mathrm{SO} (3)$ or equivalently $ \mathrm{SL} (2,  \mathbb{C} ) /  \mathrm{SU} (2)$,  which 
has an exact CFT description as a gauged WZW model. The conformal field theory  thus has level $\k$ $\mathrm{SU}(2)_\k \times \mathrm{SL}(2, \mathrm{R})_\k$ Ka\v{c}-Moody algebra as a symmetry for the left-movers and similarly for the right-movers. For superstrings in the Ramond-Neveu-Schwarz (RNS) formalism, the symmetry is enhanced to  $\mathrm{SL} (2, \mathbb{R})_\kappa \times \mathrm{SU} (2)_\kappa $ super-Ka\v{c}-Moody algebra.

The SCFT corresponding to the $T^4$ is a free-field theory whereas the SCFT corresponding to $S^3$ is given by a supersymmetric WZW model with super Ka\v{c}-Moody symmetry.  
The $\mathrm{SL} (2, \mathbb{R})_\kappa$ super Ka\v{c}-Moody algebra is described by the following OPEs,  (see,  e.g.,  \cite{Dabholkar:2007ey})
\be
\begin{aligned}
J^A (z) J^B (w) &\sim \frac{\frac{\kappa}{2} \eta^{AB} }{(z-w)^2} + \frac{\mathrm{i} f^{AB} {}_C J^C (w) }{z-w},  \\
J^A (z) \psi^B (w) &\sim \frac{\mathrm{i} f^{AB} {}_C \psi^C (w) }{z-w}, \label{alg1} \\
\psi^A (z) \psi^B (w) &\sim  \frac{\frac{\kappa}{2} \eta^{AB} }{z - w},
\end{aligned}
\ee
where  the $J^A$ are the bosonic currents and $\psi^A$ are the fermions,  both in the adjoint representation, $\eta^{AB}$ is the Killing-Cartan metric for $\mathrm{SL} (2, \mathbb{R})$ and $f^{AB} {}_C$ are the structure constants.   There is a similar anti-holomorphic copy of this algebra.  

One can define
\begin{equation} 
j^A \equiv J^A - \frac{\mathrm{i}}{\kappa} f^{A} {}_{BC} : \psi^B \psi^C :  \label{curedef}
\end{equation}
where $::$ denotes normal ordering.  The bosonic algebra is generated by $j^A$ satisfying the same OPE \eqref{alg1} but with the shifted level $\kappa \to \kappa +2$ and commutes with the fermionic algebra  generated by three free fermions. The total algebra thus factorizes. Something similar happens in case of the $\mathrm{SU}(2)_\kappa$ super Ka\v{c}-Moody algebra.   With a redefined bosonic current,  the current algebra again factorizes with a $\mathrm{SU}(2)_{\kappa-2}$ bosonic current algebra and three free fermions. In this case,  the Killing-Cartan metric is $\delta^{ab}$ and the structure constants are $\epsilon^{abc}$,  the standard Levi-Civita symbol.

\subsection{Lorentzian and Euclidean BTZ Black Hole \label{sec:BTZ}}

The Lorentzian BTZ black hole \cite{Banados:1992wn} is described by the metric
\be
\dd{s}^2 = -\frac{(r^2 - R_+^2)(r^2 -R_-^2)}{L^2 r^2} \dd{t}^2_L + \frac{L^2 r^2}{(r^2 - R_+^2)(r^2 - R_-^2)} \dd{r}^2 + r^2 \pqty{\dd{\varphi} - \frac{R_+ R_-}{L r^2} \dd{t_L}  }^2,
\label{btzmet}
\ee
where $L$ is the $\mathrm{AdS}_3$ radius,  $t_L$ is the Lorentzian time and the coordinate $\varphi$ has the range $(0,2\pi)$.  The outer and the inner horizons of the black hole are located at  $r=R_+$ and $r = R_-$ respectively.  The inverse Hawking temperature $\beta_{\rm B}$ is given by
\be
\beta_{\rm B} =  \frac{2\pi L^2 R_+}{R_+^2 - R_-^2}, \label{btzbt}
\ee
and the chemical potential is given by,
\be
\mu_{\rm B} = \frac{R_-}{L R_+}. \label{btzchemp}
\ee
It is sometimes convenient to express everything in terms of the parameters $R_+$ and $R_-$ instead of the inverse temperature $\beta_{\mathrm{B}}^{-1}$ and the chemical potential $\mu_{\mathrm{B}}$.

The Virasoro generators for boundary conformal field theory are given by
\begin{align}
L_0 - \frac{c}{24} = \frac{(R_+ - R_-)^2}{16 G L}, \quad \quad
\overline{L}_0 - \frac{\overline{c}}{24} = \frac{(R_+ +  R_-)^2}{16 G L}  \label{lob} \,,
\end{align}
where $c$ and $\bar c$ are the central charges of the holomorphic and antiholomorphic CFTs at the boundary, which for the F1-NS5 system are equal, $c = \bar c = 6 Q_1 Q_5$.  The Newton constant $G$ appearing above and in some formul\ae\ below are related to the central charge and the AdS radius by the Brown-Henneaux \cite{Brown:1986nw} formula,
\be 
G = \frac{3L}{2c} .  \label{brhe}
\ee
The mass and spin are related to the Virasoro generators by
\begin{align}
L_0 - \frac{c}{24} = \frac12( M L - J), \quad \quad
\overline{L}_0 - \frac{\overline{c}}{24} = \frac12( M L + J) , \label{lob2}
\end{align}
which imply
\begin{align}
M = \frac{R_+^2 + R_-^2}{8 G L^2}, \quad \quad
J  =  \frac{R_+ R_-}{4 G L}. \label{btzj}
\end{align}
The black hole entropy is given by the Bekenstein-Hawking formula,
\be
S = \frac{\pi R_+}{2 G}. \label{btzent}
\ee
It is easy to see that the expressions above
 satisfy the first law of thermodynamics:
\begin{equation}
\beta_{\rm B}^{-1} \, \delta S    = \delta M  - \mu_{\rm B} \,  \delta J \, .   \label{tdfl}
\end{equation}

One can analytically continue the Lorentzian time $t_L$ to the Euclidean time $t$ by ${t} = \mi t_L$ to obtain the  Euclidean BTZ metric:
\be
\dd{s}^2 = \frac{(r^2 - r_+^2)(r^2 - r_-^2)}{L^2 r^2} \dd{t}^2 + \frac{L^2 r^2}{(r^2 - r_+^2)(r^2 - r_-^2)} \dd{r}^2 + r^2 \pqty{\dd{\varphi} + \mi \frac{r_+ r_-}{L r^2} \dd{ t}  }^2. \label{btzeu}
\ee
The cross-term involving $\mi \dd{t} \dd{\varphi}$ is now imaginary. Reality of the Euclidean section can be ensured  by analytically continuing $R_\pm$ in the Lorentzian section to $r_\pm$:
\begin{align}
r_ + = R_+ \, , \qquad  r_-  = - \mi R_-  \, .
\end{align}
The boundary of Euclidean spacetime at $r = \infty$ is a conformal torus  with modular parameter given by \cite{Kraus:2006wn}
\begin{align}
\mathcal{T}_B = \frac{\mi L}{r_+ + r_-} .
\label{ttbtz}
\end{align}
One can choose a complex coordinate for the boundary torus: 
\be
2\pi w_{\rm B} = \varphi + \frac{\mi \, t}{L},\label{wbtz}
\ee
with the identifications
\be\label{ident}
 w_{\rm B}  \sim w_{\rm B}  + 1  \sim w_{\rm B} + \mathcal{T}_{\rm B}.
\ee 

\subsection[Thermal $\mathrm{AdS}_3$ and the $\mathrm{SL}(2, \mathbb{Z})$ Family]{\boldmath Thermal $\mathrm{AdS}_3$ and the $\mathrm{SL}(2, \mathbb{Z})$ Family} \label{sec:EBTZ-CFT}

For the SCFT description, it is useful to use the thermal $\mathrm{AdS}_3$ geometry with inverse temperature $\b$ and chemical potential $\m $. 
Geometrically, thermal $\mathrm{AdS}_3$ is a solid torus with the metric 
\be
 \dd{s} = \pqty{1 + \frac{{\r}^2}{L^2} } \dd{{T}} ^2 + \pqty{1 + \frac{{\r}^2}{L^2} }^{-1} \dd{{\r}} ^2 + {\r}^2 \dd{ {\theta}}^2 \, . \label{theads}
\ee
The boundary of this solid torus at $\rho \rightarrow \infty$ is a conformal torus  with coordinate $w$ and modular parameter $\mathcal{T}$ defined by\footnote{The quantities $\b$, $\mu$, $w$ and $\mathcal{T}$  refer to the thermal $\mathrm{AdS}_3$ geometry, whereas the quantities $\b_{\rm B}$, $\m_{\rm B}$, $w_{\rm B}$ and $\mathcal{T}_{\rm B}$ with the subscript ``${\rm B}$'' refer to  the BTZ black hole geometry.}

\begin{equation}
	2\pi w = \theta + \frac{\mi \, T}{L} \, , \qquad L {\cal T} =  \frac{\mi}{2\pi} \beta ( 1 + \mi L \mu)\, .  \label{wtherm}
\end{equation}
with the identifications
\begin{equation} 
	w \sim w  + 1  \sim w + \mathcal{T} \, .
\end{equation}
The path integral on this geometry with these  identifications has an interpretation as a trace over the Hilbert space $\mathcal{H}_{\rm CFT}$ of the boundary conformal field theory
\begin{equation}
	\Tr_{\mathcal{H}_{\rm CFT}} \big( \exp \left[{2\pi \mi \mathcal{T} L_0 - 2\pi \mi \overline{\mathcal{T}} \overline{L}_0}\right] \big) \, .
\end{equation}

The Euclidean BTZ black hole solution  can be related to  thermal $\mathrm{AdS}_3$ by an appropriate coordinate transformation \cite{Kraus:2006wn} but with the `space' cycle and the `time' cycle of the boundary torus exchanged by the $S$ transformation belonging to the boundary modular group ${\rm SL}(2, \mathbb{Z})$. The  boundary coordinates  and modular parameters of the two geometries are related by
\begin{equation}\label{modtrans}
{w}_{\rm B} = - \frac{w}{{ \cal T}} \, , \qquad  {\cal T}_{\rm B} = - \frac{1}{ {\cal T}} \, .
\end{equation}

In the thermal $\mathrm{AdS}_3$ geometry, the `space' cycle of the boundary torus with coordinate $\theta$  is contractible in the bulk space to a point at $\rho =0$, whereas the `time' cycle  with coordinate $T$ is non-contractible. Since the $S$ modular transformation above flips these cycles, in the BTZ geometry \eqref{btzeu} the `time' cycle of the boundary torus with coordinate $t$ is contractible in the bulk  while the `space' cycle with coordinate $\varphi$ is non-contractible. This is as expected in the Euclidean black hole geometry. The periodicity of Euclidean time at asymptotic infinity for a non-spinning black hole corresponds to the inverse temperature $\b_{\rm B}$ of the black hole, which is determined by its periodicity as an angular variable near the Euclidean horizon at the origin. 

When spacetime fermions are present in the thermal $\mathrm{AdS}_3$ geometry, they can be periodic around the non-contractible `time' cycle but must be anti-periodic at the asymptotic `space' cycle so that they are well-defined and periodic near the origin where this cycle contracts. Flipping the space and time labelings in the BTZ geometry, it corresponds to anti-periodic fermions around the time cycle, consistent with the fact the black hole has finite temperature and hence supersymmetry is broken by the boundary conditions around Euclidean time cycle.   Note that around the non-contractible cycle,  we continue to impose periodic boundary conditions on the fermions unlike in \cite{Sathiapalan:1986db,  Kogan:1987jd,  Atick:1988si}.

Instead of considering boundary tori with different modular parameters, one can fix the modular parameter and look for inequivalent classical solutions with the same asymptotics. To perform the path integral in the bulk, one must  sum over all these geometries. It is well-known \cite{Maldacena:1998bw}  that given a Euclidean boundary torus with the modular parameter ${\cal T}$, there exists an $\mathrm{SL} (2,  \mathbb{Z})$ family of solutions of the bulk Einstein equations in the interior of the solid torus. Thermal $\mathrm{AdS}_3$ and BTZ black hole are only two members of this family.  
 These inequivalent saddle points of the path integral are labeled by  entries  of an $\mathrm{SL} (2,  \mathbb{Z})$ matrix,  $a,b,c,d \in \mathbb{Z}$ with $ad  - b c =1$. The Euclidean action for each member of this family is given by \cite{Kraus:2006wn}
\be
I_{a,b,c, d} (\mathcal{T}) =  - \frac{\pi }{4G} \Im \qty(\frac{a \mathcal{T} + b}{c \mathcal{T} + d}).
\ee
The path integral is weighted by the factor $e^{-I}$,  so the dominant saddle is given by
\be
I_{\mathrm{min}} (\mathcal{T}) = \min_{ \substack{a,b,c,d \in \mathbb{Z} \\ ad - bc =1 } }  I_{a, b,c, d} (\mathcal{T}).
\ee
for a given $\mathcal{T}$. As $\mathcal{T}$ is varied, different saddle point solutions will come to dominate the path integral resulting in phase transitions. 

To illustrate the main points, we consider the special case $\mu = 0$ so that $\mathcal{T} = \mi \beta/2\pi L$ is pure imaginary. In this case, 
\be
I_{1, 0, 0, 1} (\mathcal{T}) =  - \frac{c}{12 L} \beta.
\ee
 whereas 
\be
I_{0, 1, -1, 0} (\mathcal{T}) =  - \frac{c }{12 L} \beta_{\rm B} \, ,
\ee
where $c$ is the central charge \eqref{brhe} of the boundary CFT. Note that with $\mu=0$,  $\b_{\rm B} = 4\pi^2 L^2/\b$ from \eqref{modtrans}. 
When $\b >  2 \pi L $ corresponding to low temperature, the first saddle point corresponding to the thermal $\mathrm{AdS}_3$ dominates. When $\b < 2\pi  L $ corresponding to high temperature, the second saddle point corresponding to the BTZ black hole dominates. 

The two saddles exchange dominance at the Hawking-Page \cite{Hawking:1982dh} transition  point $\beta_{\mathrm{HP}} = 2 \pi L$. In higher dimensions,  this transition has been interpreted as a confinement deconfinement transition in the boundary superconformal gauge theories \cite{Witten:1998zw}.  In two-dimensional SCFT there is no transition in the exact CFT but there can be a `phase transition' at large $c$ corresponding to the supergravity limit in the bulk.  See \cite{Maloney:2007ud} for some related comments in this context.

The world-sheet torus partition function that we evaluate in the next section computes the one-loop quantum fluctuations of the entire tower of string modes around the classical saddle point corresponding to thermal  $\mathrm{AdS}_3$ with boundary modular parameter $\mathcal{T}$. A similar calculation is valid around any other saddle point.  The one-loop partition function does not exhibit any sign of the Hawking-Page--like phase transition. This is to be expected because the phase transitions are determined by which classical action dominates the path integral, and the one-loop quantum fluctuations around a given saddle point are not sensitive to it.

\subsection[$\mathrm{AdS}_3$ World-sheet Conformal Field Theory]{\boldmath $\mathrm{AdS}_3$ World-sheet Conformal Field Theory}\label{sec:BTZ-CFT}

For the world-sheet analysis, it is convenient to consider  the coordinates \cite{Gawedzki:1991yu, Maldacena:2000kv}
\be 
\begin{aligned}
\phi &= \frac{{T}}{L} - \frac{1}{2} \log(1 + \frac{{\r}^2}{L^2} ), \,  \\
v &= \frac{{\r}}{L} e^{ \mi {\theta} }, \qquad
\bar{v} =\frac{ {\r}}{L} e^{- \mi {\theta} } \,,
\end{aligned} \label{phivco}
\ee
which brings the metric \eqref{theads} to the form suited for the path integral
\begin{align}
\dd{s}^2 = L^2 \bqty{ \dd{\phi}^2 +  ( \dd{v} + v \dd{\phi}) ( \dd{\bar v} + \bar v \dd{\phi})    }. \label{wzwmet1}
\end{align}

We would now like to describe this geometry more group-theoretically. The WZW action \cite{Gepner:1986wi} for strings on a group manifold $G$ is given by
\be
S = - \frac{\kappa}{2\pi} \int \dd[2] \sigma \, \Tr(g^{-1} \partial_\mu g g^{-1} \partial^{\mu} g ) - \frac{\mi \kappa}{24\pi} \int \dd[3] y \epsilon^{\mu \nu \rho} \Tr(g^{-1} \partial_\mu g g^{-1} \partial_\nu g g^{-1} \partial_\rho g   ), \label{wzwacx1}
\ee
where $g \in G$.
The action has  global symmetry  $G_L \times G_R$ under the transformation
\be
g(\sigma^a) \to g_L g(\sigma^a) g_R^{-1}.   \label{gtar}
\ee
We start with the six real dimensional group $ G = \mathrm{SL}(2, \mathbb{C})$ and gauge the $\mathrm{SU}(2)$ subgroup to obtain the three-dimensional Euclidean $\mathrm{AdS}_3$.  It is convenient to use in the `unitary' gauge 
the parametrization for the group element 
\be \label{gelement}
g = \begin{bmatrix}
e^{\phi} (1 + \bar{v} v) & v \\
\bar{v} & e^{-\phi} ,
\end{bmatrix} \, 
\ee
which depends now only on three real coordinates. 
The Maurer-Cartan line element
\be
\dd{s}^2 = \frac12 \Tr ( g^{-1} \dd{g} g^{-1} \dd{g})\label{maucar1}
\ee
becomes precisely a multiple of \eqref{wzwmet1}
and in the standard $(z, \bar{z})$ coordinates on the world-sheet, the action simplifies to 
\be
S \sim \int \dd[2]{z} \bqty{ \pa \phi \bar \pa \phi + (\bar \pa v + v \bar{\pa} \phi)(\pa \bar v + \bar v {\pa} \phi)  },\label{actsim2}
\ee
where $\partial \equiv \pdv*{z}$ and $\bar{\partial} \equiv \pdv*{\bar z}$.

With a non-zero chemical potential and temperature, there are additional global identifications:
\begin{align} \label{iden1} 
v \sim e^{\mi  \mu \beta} v \, , \qquad
\bar v \sim e^{-\mi  \mu \beta} \bar v \, , \qquad
\phi \sim \phi + \frac{\beta}{L} \, .
\end{align}
To implement these global  identifications on the group element, we take the ans\"{a}tze 
\begin{align}\label{ansatz}
g_L = \begin{bmatrix}
e^a & 0 \\
0 & e^{-a}
\end{bmatrix} \, , \qquad
g_R = \begin{bmatrix}
e^b & 0 \\
0 & e^{-b}
\end{bmatrix},
\end{align}
under which the group element \eqref{gelement} transforms as
\be
g \to \begin{bmatrix}
 e^{a-b+\phi }  (1+v \bar v) & e^{a+b}  v \\
 e^{-a-b} \bar v  & e^{-a+b-\phi } \\
\end{bmatrix} \label{gtrans}
\ee
This agrees with \eqref{iden1} if
\be \label{ab}
a = \frac12  \frac{\beta}{L} ( \mi \mu L + 1) = - \mi \pi {\cal T} \, , \qquad  b = \frac12 \frac{\beta}{L} ( \mi \mu L - 1) = - \mi \pi \bar{\cal T}   \, .
\ee
The modular parameter $\mathcal{T}$ was defined in \eqref{wtherm}.  Since the group action is free, this identification has no fixed points.

\section{Partition Functions of Horizon Orbifolds\label{sec:Partition}}

To construct the horizon orbifold, one must correctly identify the orbifold symmetry group $\mathbb{Z}_N$ as the subgroup of the $\mathrm{U}(1)$ group of Euclidean time translations. Note that  the thermal $\mathrm{AdS}_3$  geometry has the cycles exchanged compared to the BTZ black hole that we are interested in; the time cycle is noncontractible in the bulk whereas the space cycle is contractible. Thus, the  $\mathbb{Z}_N$  symmetry must leave the $\phi$ coordinate invariant and must shift the $\theta$ coordinate which now is the phase of the variable $v$. Its generator can be readily identified and is given by
\begin{align} \label{iden2}
v \sim e^{4\pi \mi /N} v \, , \qquad
\bar v \sim e^{-4\pi \mi /N} \bar v \, , \qquad
\phi \sim \phi \, . 
\end{align}
This can be achieved by the diagonal action $g_L = g_R$ with $a = b = 2 \pi \mi/ N$ in \eqref{ansatz}. As desired, the orbifold action has a fixed point  at the origin $v = 0$  where the bifurcate black hole horizon would be located in the BTZ geometry.  We also demand that $N$ be an \emph{odd} positive integer.  Let us mention here that $\mathbb{Z}_N$ orbifolds of $\mathrm{AdS}_3$ have been considered in the literature \cite{Martinec:2001cf,  Martinec:2002xq,  Martinec:2023zha, Gaberdiel:2023dxt} in different contexts.

\subsection{Horizon Orbifolds\label{sec:OrbPF} }

Given the identifications \eqref{iden1} and \eqref{iden2}, one would like to evaluate the path integral on the world-sheet torus with complex structure parameter $\t$ where the fields are allowed to be periodic around the cycles of the torus up to arbitrary identifications\footnote{We emphasize that the modular parameter $\t$ corresponds to the world-sheet torus in the bulk whereas the modular parameter $\mathcal{T}$ corresponds to the boundary torus.}. This implies the following boundary conditions for the fields:
\be
\begin{aligned}
\phi ( z + 2\pi) &= \phi (z) + n \frac{\beta}{L} ,  \\
\phi ( z + 2\pi \tau) &= \phi (z) + m \frac{\beta}{L},  \\
v(z + 2\pi) &= \exp( \mi n \mu \beta  + \frac{4\pi \mi}{N} k  ) v(z), \\
v(z + 2\pi \tau) &= \exp( \mi m \mu \beta  + \frac{4\pi \mi}{N} \ell  ) v (z) .
\end{aligned}
\label{phivi}
\ee
The integers $m$ and $n$ can be identified with  momentum and winding (after a Poisson resummation) around the compactified time circle whereas the integers $k$ and $\ell$ modulo $N$ correspond to the twists and twines of the $\mathbb{Z}_N$ orbifold action.

One can separate the instanton contributions by writing \cite{Maldacena:2000kv},
\begin{align}
\phi &= \hat \phi + \frac{\beta}{L} f_{m,n} (z,\bar{z}) , \\
v &= \exp( \mi \mu \beta f_{m,n} + \frac{4\pi \mi}{N} f_{\ell, k}   ) \hat{v},
\end{align}
where
\be
f_{m,n} =  \frac{m(z-\bar z) + n( \tau \bar{z} - z \bar{\tau} )}{2\pi(\tau - \bar{\tau})} \, ,
\ee
and
$\hat{\phi}$, $\hat{v}$ are periodic. The various terms in the Lagrangian effectively reduce to
\be
\pa \phi \bar \pa \phi \to \pa \hat{\phi} \bar \pa \hat{\phi}  - \frac{\beta^2}{L^2}  \frac{|m -  n \bar{\tau }|^2}{4\pi^2 (\tau - \bar{\tau})^2},
\ee
and
\be
|\bar \pa v + v \bar{\pa} \phi |^2 \to \qty| \pqty{ \bar \partial +  \bar \pa \hat{\phi} + \qty( \frac{\beta}{L} + \mi \mu \beta) {\cal U}_{m,n} + \frac{4\pi \mi }{N} {\cal U}_{k.\ell}  } \hat v  |^2 ,
\ee
where
\be
 {\cal U}_{m,n} \equiv  -\frac{m -n  \tau}{2\pi(\tau - \bar \tau)}.
\ee
The resulting action is still not quadratic but the action for the field $\hat v$ can be recognized as the action of a charged field coupled to a background gauge field. Using the chiral anomaly \cite{Polyakov:1983tt,  Polyakov:1984et} one can reduce the path integral to a Gaussian integral for $\phi$ \cite{Gawedzki:1988hq,  Gawedzki:1988nj, Gawedzki:1991yu}.
The path integral can then be readily performed to obtain the bosonic contribution to the $\mathrm{AdS}_3$ partition function for a fixed set of integers $(mn; k \ell)$.
\be
{\cal Z}^{\mathrm{AdS, B}}_{nm; k \ell} (\tau) = \frac{\beta \sqrt{\kappa}}{2\pi \sqrt{\tau_2}} \frac{\exp( - \frac{(\kappa + 2) \beta^2}{4\pi L^2 \tau_2} |m - n \tau|^2 + \frac{2\pi}{\tau_2} (\Im\ U_{nm; k\ell})^2   )}{|\vartheta_1 ( U_{nm;k\ell} | \tau) |^2} ,\label{zadsmod}
\ee
where\footnote{Note that it is $\tau$ and not $\bar{\tau}$ that appears inside the $\vartheta$-function in \eqref{zadsmod} correcting a typo in \cite{Maldacena:2000kv}. This is necessary for modular invariance as discussed in \S\ref{sec:Modular}. See also \cite{Gawedzki:1991yu}.  }
\be 
U_{nm; k\ell} \equiv (m - n \tau ) \mathcal{T} +  \frac{2}{N} (k \tau - \ell), \label{defumnkl}
\ee
We have already used $(\kappa + 2)$ for the level for this path integral for the bosonic $\mathrm{AdS}_3$ with the shift noted after eq. \eqref{curedef} to obtain the result relevant for supersymmetric $\mathrm{AdS}_3$.

It is worth noting that one can slightly simplify the notation by  using the theta function with characteristics,
\be 
\vartheta \bqty{ a \atop b } (z | \tau ) =  \sum_{n \in \mathbb{Z}} q^{  \frac{1}{2}  (n+a)^2 } e^{2 \pi \mi ( z +b) (n+ a)} \, 
\ee
and by using the fact that $\vartheta_1$ above corresponds to $a = \frac12 = b$, to obtain
\be
{\cal Z}^{\mathrm{AdS, B}}_{nm; k \ell} (\tau)= \frac{\beta \sqrt{\kappa}}{2\pi \sqrt{\tau_2}} \frac{\exp( - \frac{(\kappa +2) \beta^2}{4\pi \tau_2} |m - n \tau|^2 + \frac{2\pi}{\tau_2} (\Im\ U_{nm})^2   )}{\vqty{ \vartheta \bqty{ \frac{2k}{N} + \frac12 \atop \frac{2\ell}{N} + \frac12 } (U_{nm} | \tau) }^2 } \label{partfn2} \, ,
\ee
where $U_{nm} \equiv U_{nm;00}$. 
The $\mathbb{Z}_N$ orbifold twists and twines now appear only through the characteristics of the theta function.

\subsection[Bosons on $S^3 \times T^4$]{\boldmath Bosons on $S^3 \times T^4$}

We first note that \cite{DiFrancesco:1997nk} for the affine algebra ${\rm SU} (2)_{\kappa-2}$,  unitary representations exist only for $0 \leq \lambda \leq \kappa -2$ for integer $\lambda$ (in our convention, the ${\rm SU}(2)$ spin is $\lambda/2$).  The $S^3$ partition function is then given by the diagonal sum over characters,
\begin{equation}
{\cal Z}_{S^3} = \sum_{\lambda = 0}^{\kappa-2}  | \chi_\lambda^{(\kappa-2)} (\tau, 0 ) |^2,  \label{zs3}
\end{equation}
where the character $\chi_\lambda( \tau, z) $ is given by the Weyl-Ka\v{c} character formula,
\begin{equation}
\chi_\lambda^{(\kappa-2)} (\tau, z) =  \frac{\Theta_{\lambda+1, \kappa} (\tau,  z)- \Theta_{-\lambda-1, \kappa} (\tau, z)  }{\Theta_{1, 2} (\tau,  z)- \Theta_{-\-1, 2} (\tau,  z)  },  \label{weylkac}
\end{equation}
with $\Theta_{\lambda,  \kappa}$ being the generalized theta function
\begin{equation}
\Theta_{\lambda,  \kappa} (\tau,  z) =  \sum_{n \in \mathbb{Z} + \frac{\lambda}{2\kappa} } q^{\kappa n^2} e^{2 \pi \mathrm{i} n \kappa z}  \label{genth}.
\end{equation}
Note that $\chi_\lambda^{(\kappa-2)} (\tau, z) $ is of the indeterminate form $0/0$ for $z = 0$  and is defined by the limiting value of \eqref{weylkac} as $z \to 0$, which will be important to us later.

The torus partition function,  on the other hand,  is given by
\begin{equation}
{\cal Z}_{T^4} = \frac{1}{|\eta(\tau)|^8} \sum_{(p_R ,p_L) \in \Gamma_{4,4}} q^{p_R^2/2} \bar{q}^{p_L^2/2} , \label{zt4}
\end{equation}
where $ \Gamma_{4,4}$ is the Narain lattice \cite{Narain:1985jj,  Narain:1986am}.

\subsection[Fermions and Ghosts on $\mathrm{AdS}_3 \times S^3 \times T^4$]{\boldmath Fermions and Ghosts on $\mathrm{AdS}_3 \times S^3 \times T^4$}\label{sec:fermiads3}

Let $g(z,\bar{z})$ be an element of the group.  The components of the holomorphic current are then  given by,
\begin{equation}
J^A (z) = \kappa \Tr(T^A \partial g g^{-1} ) \label{jata},
\end{equation}
and we can also find an analogous expression for the antiholomorphic current.
A convenient set of generators in ${\rm SL} (2, \mathbb{R})$ is given by
\begin{equation}
T^3 = \frac{\mathrm{i}}{2} \sigma_z,  \quad 
\quad T^\pm = \frac12 ( \sigma_x \pm \mathrm{i} \sigma_y  ). \label{genmat}
\end{equation}
As noted in \S\ref{sec:BTZ-CFT} that the identifications  are effected by the action \eqref{gtar}
with,
\begin{align}
g_L = \mathrm{diag} (  e^{-\mathrm{i} \pi {\cal T}  },  e^{+ \mathrm{i} \pi {\cal T} }    ),\qquad
g_R = \mathrm{diag} (  e^{- \mi \pi  \overline{\cal T} },  e^{+ \mathrm{i} \pi \overline{\cal T} }    ) \label{grgliden}.
\end{align}

Under this action,   $J^3 \to J^3$, but $J^\pm$ transform as,
\begin{equation}
J^\pm \sim e^{ \pm 2 \pi \mathrm{i} {\cal T}    } J^{\pm}, \label{jpmcharge}
\end{equation}
and likewise for the antiholomorphic sector. On the other hand,  the orbifold action is effected by the group elements,
\begin{equation}
g_L = \mathrm{diag} ( e^{\frac{2\pi \mathrm{i}}{N}  },  e^{-\frac{2\pi \mathrm{i}}{N}  }    ) = g_R \label{glgrorb},
\end{equation}
which lead to similar transformation of the currents.

From these results and from \eqref{curedef}, we note that the bosonic currents $j^\pm$ and $\psi^{\pm}$ have the same charge under this action.  Therefore,  there is a single complex charged fermion and a  neutral real fermion from the $\mathrm{AdS}_3$ factor in the holomorphic sector. 
In addition, there are $7$ neutral real fermions from the $S^3$  and $T^4$ factors. Altogether, out of the five complex fermions, only one is charged with the same twist as the single charged boson consistent with world-sheet supersymmetry. The anti-holomorphic sector has the same structure. 

 A spin structure can be parametrized by $(a,b)$ on the holomorphic side and $(\bar a, \bar b)$ on the anti-holomorphic side,  where each index can take either the value 0 or 1/2. The  fermionic partition function then takes the form
 \begin{equation}
 \begin{aligned}
 {\cal Z}^{\mathrm{F}} \begin{bmatrix}
 a & \bar a \\ b & \bar{b}
\end{bmatrix}  = \frac{e^{- \frac{2\pi}{\tau_2} (\mathrm{Im} \, U_{nm; k \ell})^2   }}{|\eta(\tau)|^{10} }  & \vartheta \left[ a \atop b \right] ( U_{nm; k \ell} | \tau) \vartheta^4 \left[ a \atop b \right] ( 0 | \tau) \times \\
& \vartheta \left[ \bar a \atop \bar b \right] ( \bar{U}_{nm; k \ell} | -\bar \tau) \vartheta^4 \left[ \bar a \atop \bar b \right] ( 0 | -\bar \tau). \label{zferm}
\end{aligned}
 \end{equation}

We also have the superconformal ghosts. The bosonic ghosts do not depend on the spin structure but the fermonic ghosts do. Their contribution is given by
\begin{equation}
{\cal Z}^{\mathrm{ghost}} \begin{bmatrix}
 a & \bar a \\ b & \bar{b}
\end{bmatrix}  =  | \eta (\tau) |^4 \times \frac{ |\eta (\tau) |^2 }{ \vartheta \left[ a \atop b \right] ( 0 | \tau)   \vartheta \left[ \bar a \atop \bar b \right] ( 0 | -\bar \tau) } \times \tau_2 . \label{zgho}
\end{equation}

Thus the total part of the partition function depending on the spin structure is
\be
\begin{aligned}
\label{zfgss}
{\cal Z}^{\mathrm{F}}   Z^{\mathrm{ghost}} \begin{bmatrix}
 a & \bar a \\ b & \bar{b}
\end{bmatrix} 
 =  \frac{\tau_2 e^{- \frac{2\pi}{\tau_2} (\mathrm{Im} \, U_{nm; k \ell})^2   }}{|\eta(\tau)|^{4} }  & \vartheta \left[ a \atop b \right]  ( U_{nm; k \ell} | \tau) \vartheta^3 \left[ a \atop b \right] ( 0 | \tau) \times \\
 &\vartheta \left[ \bar a \atop \bar b \right] ( \bar{U}_{nm; k \ell} | -\tau) \vartheta^3 \left[ \bar a \atop \bar b \right] ( 0 | -\bar \tau) . 
\end{aligned}
\ee
The Gliozzi-Scherk-Olive (GSO) projection \cite{Gliozzi:1976qd} in Type-IIB theory can be implemented by taking the sum
\begin{equation}
{\cal Z}^{\mathrm{F}}   Z^{\mathrm{ghost}}= \frac14 \sum_{a,b = 0,  \frac12} \sum_{\bar{a},\bar b = 0,  \frac12} (-1)^{2a+2b} (-1)^{2\bar a + 2\bar b} {\cal Z}^{\mathrm{F}}   Z^{\mathrm{ghost}} \begin{bmatrix}
 a & \bar a \\ b & \bar{b}
\end{bmatrix}    \label{gsodef1}.
\end{equation}
We can now appeal to one of the Riemann quartic identities of theta functions (see,  e.g.,  eq. (5.8) of \cite{Dabholkar:2023yqc}) to obtain a simple form for the partition function:
\begin{equation}
\qty( {\cal Z}^{\mathrm{F}}   Z^{\mathrm{ghost}} )_{nm; k\ell}= \frac{\tau_2 e^{- \frac{2\pi}{\tau_2} (\mathrm{Im} \, U_{nm; k \ell})^2   }}{|\eta(\tau)|^{4} }  \qty| \vartheta_1 \pqty{ \frac{U_{nm; k\ell}}{2} \Big| \tau} |^8 .  \label{zfg}
\end{equation}
For earlier discussions of the superstring partition function on $\mathrm{AdS}_3\times S^3 \times T^4$,  see,   for example, \cite{Israel:2003ry,  Raju:2007uj,  Ashok:2020dnc}. 

\subsection{Superstring Partition Function}\label{sec:superfinal}

Based on the foregoing results,  we are now in a position to write down the full partition function. We take the product of  ${\cal Z}^{\mathrm{AdS, B}}_{nm; k \ell}$,  $\qty( {\cal Z}^{\mathrm{F}}   Z^{\mathrm{ghost}} )_{nm; k\ell}$  , ${\cal Z}_{S^3}$ and ${\cal Z}_{T^4}$ from eqs. \eqref{partfn2},  \eqref{zfg},  \eqref{zs3} and \eqref{zt4} respectively and then sum over $m,n \in \mathbb{Z}$ and $k,\ell \in \mathbb{Z}_N$ to obtain the full partition function,
\be
\begin{aligned}
{\cal Z} (\tau) &= \frac{\beta \sqrt{\kappa}  \sqrt{\tau_2} }{2\pi} \frac{1}{N} \frac{1}{|\eta(\tau)|^{12} }\sum_{\substack{m,n\in \mathbb{Z} \\ \ell,  k \in \mathbb{Z}_N} } \frac{e^{ - \frac{(\kappa +2) \beta^2}{4\pi  L^2 \tau_2} |m - n \tau|^2 } |\vartheta_1 ( {U}_{nm;k\ell}/2 | \tau) |^8 }{|\vartheta_1 ( {U}_{nm;k\ell} | \tau) |^2}  \\
&\quad  \times  \sum_{\lambda = 0}^{\kappa-2}  | \chi_\lambda^{(\kappa-2)} (\tau) |^2  \times \sum_{(p_R ,p_L) \in \Gamma_{4,4}} q^{p_R^2/2} \bar{q}^{p_L^2/2} . \label{final}
\end{aligned}
\ee
The factor of $1/N$ is necessary because of $\mathbb{Z}_N$ projection in the orbifold.

This is the full modular-invariant answer (see  \S\ref{sec:Modular}). The world-sheet partition function of our interest is the integral 
\be 
Z(M, J, N, \k) = \int\limits_{ \mathcal{D}} \frac{\dd[2]{\tau}}{\tau_2^2} {\cal Z} (\tau) , \label{Zmjnk}
\ee
where $ \mathcal{D}$ refers to the standard `key-hole' fundamental domain defined by $ |\tau| \geq 1$,  $|\tau_1| \leq 1/2$ and $\tau_2 > 0$.  The black hole parameters $M$ and $J$ \eqref{btzj} are related to the $\mathrm{AdS}_3$ modular parameter $\mathcal{T}$ by through the relation \eqref{modtrans},  where the BTZ modular parameter is given by \eqref{ttbtz}. We would then identify the parameter appearing in the partition function as  a rather complicated function of $M$,  $J$ and $L$ (i.e.,  $\kappa$),
\be
{\cal T} =  \frac{2 \mi \sqrt{G L \left(\sqrt{L^2 M^2-J^2}+L M\right)}}{L}  + \frac{J}{4 G L^2} \bqty{2 \sqrt{G L \left(\sqrt{L^2 M^2-J^2}+L M\right)}}^{-1}. \label{tcomp}
\ee

It is worth noting that in the final expression \eqref{final} above,  only the size of the non-contractible circle, $\beta \sqrt{\kappa}$ appears.  From the point of view of ${\rm AdS}_3$ spacetime,  this is analogous to the area of the horizon. From the ten-dimensional point of view, the sizes of the $S^3$ and $T^4$ are contained in ${\cal Z}_{S^3}$ and ${\cal Z}_{T^4}$ respectively,  as we shall see shortly.  Therefore,  we get the volume of the eight-dimensional space transverse to the two dimensional sub-manifold on which the $\mathbb{Z}_N$ acts. This is consistent with our observations in the flat space orbifolds \cite{Dabholkar:2023ows,  Dabholkar:2023yqc} . What is particularly appealing about the ${\rm AdS}_3 \times S^3 \times T^4$ geometry is that the transverse volume is naturally regulated in the infrared.
\subsection{Modular Invariance \label{sec:Modular}}

To check for modular invariance, we examine the invariance of various components of the full  partition function under the action of the generators $T$ and $S$ of the modular group  ${\rm SL}(2, \mathbb{Z})$.   The $S^3$ partition function \eqref{zs3} and the Narain partition function \eqref{zt4} are well known to be modular invariant.   It is  instructive to  analyze in some detail the modular invariance of the purely bosonic orbifold AdS partition function.  Consider
\be
{\cal Z}^{\mathrm{AdS, B}} (\tau) = \sum_{\substack{m,n\in \mathbb{Z} \\ \ell,  k \in \mathbb{Z}_N} } {\cal Z}^{\mathrm{AdS, B}}_{nm; k \ell} (\tau) , \label{zadsbn}
\ee
with ${\cal Z}^{\mathrm{AdS, B}}_{nm; k \ell} (\tau)$ given by \eqref{zadsmod}.  Under the $T$ transformation, $\tau \to \tau + 1$,  we find
\begin{align}
m - n \tau  \to (m -n) - n \tau, \qquad U_{nm; k\ell} \to U_{n(m-n); k(\ell - k)} . \label{trta}
\end{align}
Invariance of \eqref{zadsbn} under  $T$-transformation is then easily checked  by shifting the $m,  \ell$ sums appropriately.

Under the $S$  transformation,  $\tau \to - 1 / \tau$, we have
\be
U_{nm; k\ell} \to  -\frac{1}{\tau} U_{m(-n); \ell(-k)}   \label{stranu}.
\ee
The Jacobi theta function and the Dedekind eta function transform as
\be
\begin{aligned}
 \vartheta_1 \qty(-\frac{z}{\tau} \Big| - \frac{1}{\tau}) &=  \sqrt{-\mi \tau} \exp( \frac{ \mi \pi z^2}{ \tau}  ) \vartheta_1 (z| \tau),  \\
\eta \pqty{ - \frac{1}{\tau}} &= \sqrt{-\mi \tau} \eta(\tau)
  \label{jacotra}
\end{aligned}
\ee
respectively.

Using these relations one finds after some tedious algebra,
\begin{align}
{\cal Z}^{\mathrm{AdS, B}} (-1/\tau) &= \frac{\beta \sqrt{\kappa}}{2\pi \sqrt{\tau_2}} \frac{1}{N}\sum_{\substack{m,n\in \mathbb{Z} \\ \ell,  k \in \mathbb{Z}_N} } \frac{\exp( - \frac{(\kappa + 2)\beta^2}{4\pi L^2 \tau_2} |n - m \tau|^2 + \frac{2\pi}{\tau_2} (\Im\ U_{m(-n); \ell(-k)})^2   )}{|\vartheta_1 ( U_{m(-n); \ell(-k)}| \tau) |^2} \non \\
&= {\cal Z}^{\mathrm{AdS, B}}  (\tau) . \label{modinv}
\end{align}
This proves that the  bosonic partition function is invariant under the full modular group\footnote{The partition function after the $\tau$ integrations should be invariant also under the modular transformations of the boundary modular parameter $\mathcal{T}$. It would be interesting to exhibit this modular invariance explicitly but this will not be our concern here.}.  Putting all these elements together,  it is easy to show that the full partition function including the fermions and ghosts
\be
{\cal Z}^{\mathrm{AdS}} (\tau) = \sum_{\substack{m,n\in \mathbb{Z} \\ \ell,  k \in \mathbb{Z}_N} } {\cal Z}^{\mathrm{AdS, B}}_{nm; k \ell} (\tau) \qty( {\cal Z}^{\mathrm{F}}   Z^{\mathrm{ghost}} )_{nm; k\ell} , \label{zadsfn}
\ee 
is also modular invariant even though individual terms of the sum are not.

\section{Analytic Continuation and Entanglement Entropy \label{sec:Ent}}

 We would like to identify  all tachyonic terms in the partition function \eqref{final} that grow exponentially as $\tau_2$ becomes large.   Our goal is to examine if the total tachyonic contribution can be resummed and analytically continued to the physical region $0 < N \leq 1$ \cite{Dabholkar:2023ows,  Dabholkar:2023yqc}.  For computing the von Neumann entropy,  we need the derivative of the analytically continued partition function with respect to $N$ at $N=1$.
 
We are interested in relating our results to those of asymptotically flat spacetime.  We had previously mentioned in \S\ref{AdS3-CFT}  that the common radius of curvature of $\mathrm{AdS}_3$ and the $S^3$ is given by,
\be
L =\sqrt{\kappa} \ell_s.  \label{lkappals} 
\ee
Therefore,  the flat space limit of the $\mathrm{AdS}_3$ and the $S^3$ (in which case,  it is the decompactification limit) is achieved by taking
\be 
\kappa \to \infty .  \label{kappainf} 
\ee

The flat space limit can be taken in two inequivalent ways:  we can keep $\beta$ fixed or scale $\beta$ with $\kappa$ in a suitable way. We recall from the discussion at the end of \S\ref{sec:superfinal} that the non-contractible circle of ${\rm AdS}_3$ has the size $\beta \sqrt{\kappa}$. If we keep $\beta/L$ fixed and take the limit \eqref{kappainf},  the circle becomes of infinite size.  This limit is related to the situation explored in \cite{Dabholkar:2023ows,  Dabholkar:2023yqc}.

The other interesting limit is when we take $\beta/L \to 0$ with \eqref{kappainf} in a way so that,
\be
\frac{\beta \sqrt{\kappa}}{L} = \text{finite}. \label{otherscal} 
\ee
In this scaling, the size of the circle remains finite even in the flat-space limit.

\subsection[Tachyons in $\mathrm{AdS}_3$ at Finite Curvature]{\boldmath Tachyons in $\mathrm{AdS}_3$ at Finite Curvature}

We now consider the partition function \eqref{final}. 
To simplify the analysis it is useful to use the unfolding method  \cite{Polchinski:1985zf} and trade the integrand with the  $n$-sum integrated over the key-hole fundamental domain $\mathcal{D}$ (defined below eq.  \eqref{Zmjnk})   for an integrand without the $n$ sum (with $n$ set to 0) with the $m$-sum running over all non-positive integers. 
 The function\footnote{The $(m,n) = (0,0)$ term is modular invariant by itself and hence the region of integration continues to be the fundamental domain $\mathcal{D}$.  In the following analysis, we discuss all the $m=0$ and $m\neq 0$ terms together since the $\tau_2 \to \infty$ region is common for all the terms.} now has to be integrated over the strip $|\tau_1| \leq 1$,  $\tau_2 >0$.  

We are interested the ${\rm AdS}_3$  part of the partition function,  which after unfolding takes the form
\be
 {\cal Z}^{\rm A} (\tau) = \frac{1}{|\eta(\tau)|^{18} }  \frac{1}{N}\sum_{\substack{m \in \mathbb{Z}_{\geq 0} \\ \ell,  k \in \mathbb{Z}_N} } \frac{e^{ - \frac{(\kappa +2) \beta^2}{4\pi L^2 \tau_2} m^2 } |\vartheta_1 \pqty{ \frac{{U}_{0(-m);k\ell}}{2} | \tau}  |^8 }{|\vartheta_1 ( {U}_{0(-m);k\ell} | \tau) |^2}. \label{zatau}
\ee
We have put the Dedekind eta functions (some of which come from other parts of the partition function) to facilitate comparison with the flat-space answer \cite{Dabholkar:2023ows}.  The remaining part of the partition function,  which we will deal with in the next section,  has the form,
\be 
 {\cal Z}^{\rm R} (\tau) =  \frac{\beta \sqrt{\kappa}  \sqrt{\tau_2} }{2\pi} |\eta(\tau)|^6  \sum_{\lambda = 0}^{\kappa-2}  | \chi_\lambda^{(\kappa-2)} (\tau) |^2   \sum_{(p_R ,p_L) \in \Gamma_{4,4}} q^{p_R^2/2} \bar{q}^{p_L^2/2}.  \label{remain}
\ee
Therefore,  the total function \eqref{final} has the form ${\cal Z} (\tau) =  {\cal Z}^{\rm A} (\tau)  {\cal Z}^{\rm R} (\tau) $.
We now look at the sum \eqref{zatau} term by term in $m$.

Let us first examine the flat space limit \eqref{kappainf} with fixed $\beta$.  In this case,  only the $m = 0$ term in \eqref{zatau} survives.  This yields
\be
 {\cal Z}^{\rm A} (\tau) = \frac{1}{N}\sum_{  k,  \ell \in \mathbb{Z}_N} \left| \frac{\vartheta_1^4 \pqty{ \frac{k\tau - \ell}{N}  | \tau} }{\eta^9 (\tau) \vartheta_1 \pqty{ \frac{2k\tau - 2\ell}{N}  | \tau}} \right|^2. \label{zaflasp}
\ee
This is precisely the combination of $\vartheta$ and $\eta$ functions that appears in the calculation of entanglement entropy \cite{Dabholkar:2023ows} associated with Rindler observers. 

We now proceed to deal with the partition function at finite $\kappa$.  Let us first find the leading term in a large $\tau_2$ expansion.  For simplicity,  we set $\mu = 0$. The  argument of the theta function then becomes
\be
{U}_{0(-m);k\ell} = -m \mi \frac{\beta}{2\pi L} + \frac{2}{N} ( k \tau - \ell). 
\ee
We would first like to find the ground state energy from the asymptotics of the theta function. 
The presence of the $-m \mi \beta$ term can be interpreted as a thermal twisting.  In the flat-space limit, the leading tachyonic term in each $k$ sector goes like \cite{Dabholkar:2023ows,  Dabholkar:2023yqc}
\be
\exp( \frac{2k}{N} 2\pi \tau_2  ) .
\ee
In the modified case,  it would therefore go like,
\be
\exp(-m \frac{\beta}{L} )  \exp( \frac{2k}{N} 2\pi \tau_2  ) .
\ee
However,  for the sub-leading tachyons, the contributions are known to come from the twisted bosonic oscillators \cite{Dabholkar:2023ows,  Dabholkar:2023yqc} corresponding to the theta function in the denominator.  These give a contribution as follows:
\be
\exp(-m \frac{\beta}{L} ) \exp( \frac{2k}{N} 2\pi \tau_2  )  \sum_{n = 0}^{n_k} \exp( - 2mn \frac{\beta}{L}  -  4 \pi \tau_2 n  \pqty{1 - \frac{2k}{N} } ),
\ee
where $n_k$ is a non-negative integer so that the sum includes only tachyons, i.e.,  terms that grow exponentially with $\tau_2$.  In our proposed method \cite{Dabholkar:2023ows, Dabholkar:2023yqc} of including all the tachyons,  one takes  $n_k \to \infty$ and performs the $k$-sum in the twisted sectors.  This method yields the contribution,
\be 
\widetilde{\mathcal{F}}_{0,m}(\tau, \mathcal{T} ) = -2 \sum_{n=0}^\infty \exp (- 4 n \pi \tau_2 - \pqty{2n + 1}m  \frac{\beta}{L} ) \frac{1-\exp(\frac{ {(N-1) }  (2 n+1)2 \pi \tau _2}{N})}{1-\exp(-\frac{(2 n +1) 4 \pi   \tau _2}{N})}.  \label{ftil1}
\ee
This is the required formula for the resummed tachyonic subtraction in AdS$_3$ for each $m$ valid for any value of complex $\mathcal{T}$.  Note that the tachyonic subtraction does not involve the angular potential $\mu$ and only the inverse temperature enters.  This fact can be understood as follows.  The sum over the twines $\ell$ in the orbifold enforces  projection operation and only $\mathbb{Z}_N$-invariant states contribute.  It turns out that as in \cite{Dabholkar:2023ows,  Dabholkar:2023yqc},   $\mathbb{Z}_N$-invariance automatically enforces level-matching as far the leading and subleading tachyons are concerned. Thus, after the sum over the twines, the tachyonic integrand automatically becomes independent of $\Re \, \tau$.  The total tachyonic contribution to be subtracted is therefore given by
\be
\begin{aligned}
 \widetilde{\mathcal{F}}_{0}(\tau,  \mathcal{T} ) &= \sum_{m=0}^\infty  e^{ - \frac{(\kappa +2) \beta^2}{4\pi L^2 \tau_2} m^2 } \widetilde{\mathcal{F}}_{0,m}(\tau,   \mathcal{T}) \\
&=  -2 \sum_{m=0}^\infty  e^{ - \frac{(\kappa +2) \beta^2}{4\p L^2 \tau_2} m^2 }  \sum_{n=0}^\infty \e^{- 4 n \pi \tau_2 - \pqty{2n + 1}m  \frac{\beta}{L} } \frac{1-\exp(\frac{ {(N-1) }  (2 n+1)2 \pi \tau _2}{N})}{1-\exp(-\frac{(2 n +1) 4 \pi   \tau _2}{N})} .
\end{aligned} \label{central}
\ee
This is an exact formula which continues in a nice manner to $0 < N \leq 1$ --- it remains finite as $\tau_2 \to \infty$.  This expression is applicable for all values\footnote{ We  assume that value of $\kappa$ is sufficiently large so that the temperature of the black hole lies below the Hagedorn temperarture.   Near this temperature,  one expects additional divergences corresponding to the Hagedorn transition   \cite{Lin:2007gi, Berkooz:2007fe, Rangamani:2007fz},  which are not the focus of  the present investigation}  of $\kappa$ and $\beta$.  In particular,  since the BTZ black hole is the dominant configuration in the thermal ensemble for small enough values of $\beta$, this tachyonic contribution also applies to the BTZ black hole in the appropriate regime.   We have to multiply $ \widetilde{\mathcal{F}}_{0}(\tau,  \mathcal{T} )$ with the remaining part ${\cal Z}^{\rm R} (\tau)$ in \eqref{remain}. We can subtract this contribution from the full partition function \eqref{final}. The remainder\footnote{There are ``long string''  divergences from the interior of the moduli space which we discuss later. } is free of tachyonic divergences from the large $\tau_2$ region.  This is the central result of our paper.

\subsection{\boldmath Flat-Space Limit and Finite Curvature Corrections}

Let us discuss the other contributions to the partition function \eqref{remain}.  Let us first look at the contribution from the Narain partition function in the limit of large volume.  In this limit, we will obtain non-compact momentum integral for $T^4$,
\begin{equation}
\sum_{(p_R ,p_L) \in \Gamma_{4,4}} q^{p_R^2/2} \bar{q}^{p_L^2/2}  \to \frac{V_4 }{\ell_s^4 (\sqrt{\tau_2})^4} = \frac{V_{4}}{\ell_s^4  \tau_2^2}. \label{momint}
\end{equation}
Note that we are using a normalization in which the volume of the torus is $(2\pi)^4 V_4$.  We could consider finite corrections to this approximation --- these will be exponentially suppressed in the volume of the torus.

Let us now consider the $S^3$ partition function.
As noted previously, that the character $\chi_\lambda^{(\kappa-2)} (\tau, z) $ is of the indeterminate form $0/0$ for $z = 0$.  Therefore,  in order to evaluate this quantity,  we should take the limiting value of \eqref{weylkac} as $z \to 0$.  We can thus write, after some calculations and simplifications,
\be
|\eta(\tau)|^6  \sum_{\lambda = 0}^{\kappa-2}  | \chi_\lambda^{(\kappa-2)} (\tau) |^2=    \sum_{\lambda = 0}^{\kappa-2}   \qty| \sum_{n \in \mathbb{Z} } (\lambda +2 \kappa  n+1) q^{\frac{(\lambda +2 \kappa  n+1)^2}{4 \kappa }}  |^2.  \label{zs3simp}
\ee
In the $\kappa \to \infty$ limit,  to obtain the leading term in the analysis (and indeed any terms  suppressed by powers of $\kappa$ in the sum),  we have to look at only $n = 0$ term in the above sum. Any other term will involve exponentially small terms of the form
$$
e^{ - 4 \pi \tau_2  n^2 \kappa  }. 
$$
These terms cannot yield corrections which are power-law in $\kappa$ (in the integrand). Therefore,  the leading answer should come from just the $n=0$ term and the sum over spins $\lambda$.  We simply have to do the following Riemann sum:
\be
\begin{aligned}
\sum_{\lambda = 0}^{\kappa -2} (\lambda + 1)^2 e^{ - \pi \tau_2 \frac{(\lambda + 1)^2}{\kappa}  }  &=  \kappa^{3/2}  \frac{1}{\sqrt{\kappa}} \sum_{\lambda = 0}^{(\sqrt{\kappa} - \frac{2}{\sqrt{\kappa}}) \sqrt{\kappa} } \qty( \frac{\lambda + 1}{\sqrt{\kappa}})^2 e^{ - \pi \tau_2 \frac{(\lambda + 1)^2}{\kappa}  }  \\
&\to \kappa^{3/2} \int_0^\infty x^2 e^{ - \pi \tau_2 x^2  } \\
&= \frac{1}{4\pi} \pqty{  \frac{\kappa}{\tau_2}  }^{3/2}. \label{s3riem}
\end{aligned}
\ee
We can further refine this calculation,  by making a careful use of the Euler-Maclaurin sum formula to deduce the possibly power-law corrections to the formula above. The Euler-Maclaurin sum formula reads,
\be 
\begin{aligned}
\sum_{j = 0}^n f(a + j h) &= \frac{1}{h} \int\limits_{a}^{a + nh } f(x) \dd{x} + \frac{1}{2} \bqty{ f(a) + f(a+nh) } \\
&\quad + \sum_{k = 1}^m \frac{B_{2k}}{(2k)!} h^{2k -1} \bqty{ f^{(2k-1)} (a+ n h) - f^{(2k-1)} (a)  } + R_m, \label{eulma}
\end{aligned}
\ee
where $R_m$ is a remainder term and $B_{2k}$ are the Bernoulli numbers.

Using the functional forms and setting the limits appropriately,  we find a surprising result: when we add up all the contributions,  the potential power-law corrections in $\kappa$ sum up to zero.  The corrections are exponentially small in $\kappa$:
\be 
\sum_{\lambda = 0}^{\kappa -2} (\lambda + 1)^2 e^{ - \pi \tau_2 \frac{(\lambda + 1)^2}{\kappa}  }  =  \frac{1}{4\pi} \pqty{  \frac{\kappa}{\tau_2}  }^{3/2}  + \mathcal{O} (e^{ - a(\tau_2) \kappa } ),  \label{s3expc}
\ee
where $a(\tau)$ is some positive constant.  One can check numerically that such a non-trivial cancellation indeed takes place and the constant $a(\tau)$ for different fixed values of $\tau$ can also be determined. 

Let us now make some comments on the total partition function in the flat limit \eqref{kappainf} with a constant $\beta$.  We also consider the large torus limit.  From \eqref{zaflasp},  \eqref{momint} and \eqref{s3expc},  we find the leading term in a large-$\kappa$ expansion to be
\be
\frac{\beta \kappa^2 V_{4} }{8\pi^2  \ell_s^4 \tau_2^3} \frac{1}{N}\sum_{  k,  \ell \in \mathbb{Z}_N} \left| \frac{\vartheta_1^4 \pqty{ \frac{k\tau - \ell}{N}  | \tau} }{\eta^9 (\tau) \vartheta_1 \pqty{ \frac{2k\tau - 2\ell}{N}  | \tau}} \right|^2.
\ee
This is precisely the flat space contribution.  In a large $\kappa$ expansion,  the remaining terms in the integrand are exponentially suppressed in $\kappa$.  For example,  the contribution from the ${\rm AdS}$ part would be suppressed by factors of $\exp(-(\kappa+2) \beta^2 / 4\pi L^2 \tau_2)$,  see \eqref{zatau}.  The finite-$\kappa$ contributions from $S^3$ are also exponentially suppressed in $\kappa$,  see eq. \eqref{s3expc}.
Thus,  all corrections,  both from AdS$_3$ and $S^3$, are exponentially small in $\kappa$ in the integrand.  These can lead to possible power-law corrections \emph{after} the modular integration.

One can draw two main lessons from this analysis of the flat space limit. 
\begin{itemize}
\item The entanglement entropy that we are computing is localized  and is proportional to the area of the entangling surface and not to the bulk volume as would be the case for the entropy of the thermal bath surrounding the black hole. 
\item There are no power-law corrections in $\kappa$ in the integrand,  as one would expect for a fluffy object like a string straddling across the entangling surface with infinite tower of massive string states.   Possible power-law corrections from the integral would be analogous to the Wald entropy \cite{Wald:1993nt} coming from higher derivative terms in the local Lagrangian.  
\end{itemize}

\subsection{Relation to Thermal Entropy and Holography \label{sec:relation}}

We have been able to provide a reasonable definition of entanglement entropy in string theory for an entangling surface that coincides with the black hole horizon, and find an  expression for this quantity that is finite both in the ultraviolet and the infrared.   The resulting entropy has non-trivial dependence on $M$, $J$ and $\k$ which would be interesting to analyze further. An additional advantage of computing the entropy for this string background is that it allows us to place the question of entanglement entropy in quantum gravity within the context of  holography where one can make contact with the boundary theory. Much is known on both sides of the $\mathrm{AdS}_3/\mathrm{CFT}_2$ correspondence,  especially in the tensionless limit  (which corresponds to $\kappa = 1$, the minimal value of the three-form flux) and one  can hope to learn more about the physical meaning of the quantity that we are defining. It is worth exploring the corresponding computation in the dual theory using recent results for the dual description \cite{Gaberdiel:2018rqv,  Giribet:2018ada,  Eberhardt:2018ouy,  Eberhardt:2019ywk,  Eberhardt:2020bgq,  Dei:2020zui, Gaberdiel:2021kkp, Gaberdiel:2023dxt}.  It would also be interesting to relate our results to matrix model calculations \cite{Hartnoll:2015fca,  Das:2020jhy, Das:2020xoa},  where the ultraviolet cut-off is the Planck scale rather than the string scale.

The entanglement entropy that we have defined is proportional to the area of the entangling surface. This is clear from the powers of $\k$ in the large $\k$ flat-space limit considered above. Therefore, it is quite distinct from the thermal entropy of the heat bath surrounding the black hole which would be proportional to the volume of bulk space with an additional power of $\kappa$.  It thus appears to be a quantity that is essentially localized at the entangling surface. The orbifold method allows one to focus on the bifurcate horizon which is the entangling surface for the inside and outside of the black hole and which in the Euclidean continuation corresponds to the orbifold fixed point. Since the quantity that we have constructed receives contributions only from the twisted sectors localized at the orbifold fixed point, it is proportional to the area of the entangling surface. 

It is worth recalling the interpretation of the thermal entropy from the boundary perspective summarized in \cite{Dabholkar:2022mxo, Dabholkar:2023ows}. The two-sided bulk black hole  geometry has two boundaries with two copies of left and right conformal field theories of states living on the left and the right boundaries respectively. 
In a two-sided BTZ black hole geometry, the full boundary Hilbert space is a direct product of the Hilbert spaces corresponding to the CFTs living on the two asymptotic boundaries.  The Hartle-Hawking vacuum state $| \Omega \rangle$ in the bulk corresponds to the maximally entangled thermofield double pure state \cite{Maldacena:2001kr, VanRaamsdonk:2010pw,Maldacena:2013xja} in the product Hilbert space.  After taking a partial trace of this pure state over the left Hilbert space,  one obtains a thermal density matrix on the right, corresponding to the thermal nature of the black hole as seen by an observer on the right.
The entropy of the thermal bath in the black hole geometry corresponds to the entropy of the thermal density matrix in the boundary theory. In the field theory approximation of string theory, this quantity is ultraviolet divergent even though there is no corresponding divergence in the boundary theory.  One of the most fruitful directions of exploration would be the relation between our results  and holographic entanglement entropy and related notions \cite{Ryu:2006bv, Hubeny:2007xt,  Faulkner:2013ana,  Jafferis:2015del}.

It is interesting to explore how the entanglement entropy defined by the orbifold method is related to the thermal entropy of the gas of particles around a BTZ black hole. It was pointed out in \cite{Sun:2020ame} that the one-loop partition function of a quantum field in ${\rm AdS}$ spacetime has a natural decomposition in terms of bulk and edge characters. The same is expected to be true even after the global identifications that define BTZ black hole. A natural conjecture would be that the entanglement entropy defined here is focusing on the contributions of the edge modes. 
  
It would also be worthwhile to explore the spectrum of the theory after the analytic continuation,  i.e., in the region $0 < N \leq 1$.  The spectrum of physical states in ${\rm AdS}_3$ was analyzed in great detail in \cite{Maldacena:2000hw}.  It was observed that in addition to the standard representations of ${\rm SL} (2,  \mathbb{R})$,  one must consider additional representations generated by the spectral flow operation. These considerations are important for the ``long string'' states in the spectrum.  It was shown in \cite{Maldacena:2000kv} that the full spectrum, including the spectrally flowed states,  can be read off from the modular invariant partition function.  The long string states are associated with poles of the partition function in the region of integration on the complex $\tau$ plane.  For the partition  function \eqref{zatau} of our consideration,  these poles appear at values of $\tau$ satisfying $U_{0(-m); k\ell}= -w \tau -s$, for odd integral values of $w$ and $s$ (there would be no poles for even values $w$ or $s$ because of the zero from the fermionic numerator).  These poles would be located at
\be
\tau =  \frac{2 \ell -N s +m N {\cal T}}{2 k+N w}. \label{taupole}
\ee
The authors of \cite{Maldacena:2000kv} argued that the divergences from these poles are actually associated with the infinite volume of ${\rm AdS}_3$ spacetime.  These long string divergences do not represent an instability like the tachyons from $\tau_2 \to \infty$ do.  Furthermore,  these occur only at non-zero values of $m$ and therefore do not occur in the leading order term \eqref{zaflasp}.  In a large-$\kappa$ expansion,  these terms are exponentially suppressed in $\kappa$ in the integrand,  as argued previously. It would be interesting to explore the fate of the long strings after the analytic continuation.
  
The quantity that we have defined is a stringy generalization of the ultraviolet divergent quantity defined by replica method in quantum field theory limit. It is satisfying that in string theory this quantity appears to be finite  and possibly computable explicitly as a function of various parameters. This should have important implications for the black hole information paradox. 

We leave these tantalizing questions for future explorations.

\section*{Acknowledgements}

We thank Dionysios Anninos,  Sumit R.  Das,  Andrea Dei,  Kyriakos Papadodimas, Alfred D. Shapere  and Edward Witten for useful discussions.  U.M. thanks the University of Chicago,  the University of Kentucky, Stony Brook University and Harvard University for their warm hospitality while this work was being completed.

\bibliographystyle{JHEP}
\bibliography{entangle.bib}

\end{document}